\documentclass[%
 reprint,
superscriptaddress,
 amsmath,amssymb,
 aps,
 pra,
floatfix,
]{revtex4-2}

\usepackage{graphicx}
\usepackage{dcolumn}
\usepackage{bm}
\usepackage[colorlinks=true,allcolors=blue]{hyperref}
\usepackage{siunitx}
\usepackage{multirow}
\newcommand{\eref}[1]{Eq.~\ref{#1}}
\newcommand{\fref}[1]{Fig.~\ref{#1}}
\newcommand{\tref}[1]{Tab.~\ref{#1}}

\newcommand{\SM}{Appendix}
\newcommand{\InGaPx}{$\textrm{In}_{1-x}\textrm{Ga}_x\textrm{P}$}
\newcommand{\InGaPval}{$\textrm{In}_{0.43}\textrm{Ga}_{0.57}\textrm{P}$}

\begin{document}

\preprint{APS/123-QED}

\title{High-Q trampoline resonators from strained crystalline InGaP for integrated free-space optomechanics}


\author{Sushanth Kini Manjeshwar}
\affiliation{Department of Microtechnology and Nanoscience (MC2), Chalmers University of Technology, SE-412 96 Gothenburg, Sweden}

\author{Anastasiia Ciers}
\affiliation{Department of Microtechnology and Nanoscience (MC2), Chalmers University of Technology, SE-412 96 Gothenburg, Sweden}

\author{Fia Hellman}
\affiliation{Department of Microtechnology and Nanoscience (MC2), Chalmers University of Technology, SE-412 96 Gothenburg, Sweden}

\author{J{\"u}rgen B{l\"a}sing}
\affiliation{Institute of Physics, Otto von Guericke Universität Magdeburg, 39106 Magdeburg, Germany}

\author{André Strittmater}
\affiliation{Institute of Physics, Otto von Guericke Universität Magdeburg, 39106 Magdeburg, Germany}

\author{Witlef Wieczorek}
\email{witlef.wieczorek@chalmers.se} 
\affiliation{Department of Microtechnology and Nanoscience (MC2), Chalmers University of Technology, SE-412 96 Gothenburg, Sweden}

\date{\today}

\begin{abstract}
Tensile-strained materials have been used to fabricate nano- and micromechanical resonators with ultra-low mechanical dissipation in the kHz to MHz frequency range. These mechanical resonators are of particular interest for force sensing applications and quantum optomechanics at room temperature. Tensile-strained crystalline materials that are compatible with epitaxial growth of heterostructures would thereby allow realizing monolithic free-space optomechanical devices, which benefit from stability, ultra-small mode volumes, and scalability. In our work, we demonstrate string- and trampoline resonators made from tensile-strained InGaP, which is a crystalline material that can be epitaxially grown on an AlGaAs heterostructure.  The strain of the InGaP layer is defined via its Ga content when grown on (Al,Ga)As. In our case, we realize devices with a stress of up to 470\,MPa along the $[1\,1\,0]$ crystal direction. We characterize the mechanical properties of the suspended InGaP devices, such as anisotropic stress, yield strength, and intrinsic quality factor. We find that the latter degrades over time. We reach mechanical quality factors surpassing $10^7$ at room temperature with a $Q\cdot f$-product as high as $7\cdot10^{11}\,$Hz with trampoline-shaped micromechanical resonators, which exploit strain engineering to dilute mechanical dissipation. The large area of the suspended trampoline resonator allows us to pattern a photonic crystal to engineer its out-of-plane reflectivity in the telecom band, which is desired for efficient signal transduction of mechanical motion to light. Stabilization of the intrinsic quality factor together with a further reduction of mechanical dissipation through hierarchical clamping or machine learning-based optimization methods paves the way for integrated free-space quantum optomechanics at room temperature in a crystalline material platform.
\end{abstract}

\maketitle

\section{Introduction}

Mechanical dissipation in nano- and micromechanical resonators has been drastically reduced in recent years by the use of dissipation dilution, soft clamping, and strain-engineering techniques \cite{gonzalez1994brownian,unterreithmeierDampingNanomechanicalResonators2010,schmidDampingMechanismsHigh2011,tsaturyan2017ultracoherent,ghadimi2018elastic,sementilli_nanomechanical_2022}. Most of these methods require the use of tensile-strained materials, such as the widely employed amorphous SiN \cite{verbridgeHighQualityFactor2006,norte2016mechanical,reinhardtUltralowNoiseSiNTrampoline2016,tsaturyan2017ultracoherent,ghadimi2018elastic,reetzAnalysisMembranePhononic2019,hojUltracoherentNanomechanicalResonators2021,shin2022spiderweb} and, more recently, crystalline materials such as SiC \cite{kermany2014microresonators,romero2020engineering}, Si \cite{beccari2022strained}, GaNAs \cite{onomitsuUltrahighQMicromechanicalResonators2013} and InGaP \cite{Cole2014,Buckle2018,Buckle2021}. Ultrahigh-quality-factor mechanical resonators fabricated from these materials open up exciting prospects for nanomechanical sensing by reaching unprecedented force sensitivities \cite{masonContinuousForceDisplacement2019,halgMembraneBasedScanningForce2021,beccari2022strained} and, when the resonators are coupled to light, pave the way for generating optomechanical quantum states at room-temperature \cite{norte2016mechanical,maccabe2020nano,barzanjehOptomechanicsQuantumTechnologies2022}. 

Strained crystalline materials compatible with epitaxial layer growth can realize integrated cavity optomechanical devices through bottom-up growth and top-down microfabrication. At the same time, this integrated approach would enable on-chip stability and scalability. Current optomechanical devices incorporating chip-based mechanical resonators that are coupled to out-of-plane light resort to stacking of multiple chips \cite{nair2017optomechanical} or to assembling independent components \cite{piergentiliTwomembraneCavityOptomechanics2018,gartner2018integrated}. Integrating the free-space optical cavity and the mechanical resonator on a single chip would provide alignment-free devices with ultra-small mode volumes to drastically increase the interaction strength between out-of-plane light and mechanical motion.

InGaP is a crystalline material that can be epitaxially grown on (Al,Ga)As and, therefore, would enable realization of integrated free-space cavity optomechanics on a chip. Further, InGaP can be grown with tensile strain on (Al,Ga)As determined by the Ga content of the InGaP layer, and, thus, has the potential to lead to ultra-low dissipation mechanical resonators. Tensile-strained micromechanical resonators fabricated from InGaP have been recently demonstrated in membrane \cite{Cole2014} and string-type geometries \cite{Buckle2018,Buckle2021}. Membrane-type micromechanical resonators have thereby reached quality factors of up to $10^6$ at room temperature \cite{Cole2014}. Further, it was experimentally confirmed that stress is anisotropic in InGaP \cite{Buckle2018}, which opens up new avenues for strain engineering the geometry of nano- and micromechanical resonators.

In our work, we demonstrate trampoline-shaped micromechanical InGaP resonators that combine low mechanical dissipation with engineered optical reflectivity, a crucial step towards free-space cavity optomechanics on a chip. We achieve mechanical quality factors surpassing $10^7$ at room temperature with trampoline-shaped micromechanical resonators, which employ a simple geometry to dilute the material’s intrinsic dissipation \cite{norte2016mechanical,reinhardtUltralowNoiseSiNTrampoline2016,romero2020engineering,bereyhi2022hierarchical}. For transduction of mechanical displacement to the light field we engineer the out-of-plane reflectivity of the resonator at telecom wavelengths by patterning the central area of the \SI{73}{\nano\meter}-thick InGaP trampoline with a photonic crystal  \cite{buiHighreflectivityHighQMicromechanical2012,makles2DPhotoniccrystalOptomechanical2015,norte2016mechanical,bernardPrecisionResonanceTuning2016,gartner2018integrated,Kini2020}. We first study the mechanical material properties of the strained InGaP layer \cite{Cole2014,Buckle2018,Buckle2021} by fabricating and characterizing string resonators to determine the intrinsic stress, yield strength, and the intrinsic quality factor. We then demonstrate high-$Q$ trampoline-shaped InGaP micromechanical resonators with engineered optical reflectivity.

\section{Fabrication}

We fabricate InGaP string- and trampoline-shaped mechanical resonators from a III-V material heterostructure grown via metal-organic chemical vapor deposition (MOCVD). A \SI{400}{\nano\meter}-thick GaAs buffer layer is grown on a GaAs substrate along the $[0\,0\,1]$ crystal direction followed by a \SI{73}{\nano\meter}-thick \InGaPx{} layer, a sacrificial layer of $\textrm{Al}_{0.68}\textrm{Ga}_{0.32}\textrm{As}$  with a thickness of \SI{785}{\nano\meter}, and another \InGaPx{} layer of \SI{75}{\nano\meter}-thickness. Note that the as-grown structure would allow for implementing sub-\SI{}{\micro\meter}-spaced two-element optomechanics on a chip \cite{gartner2018integrated}. 

\begin{figure}[t!hbp]
    \centering\includegraphics{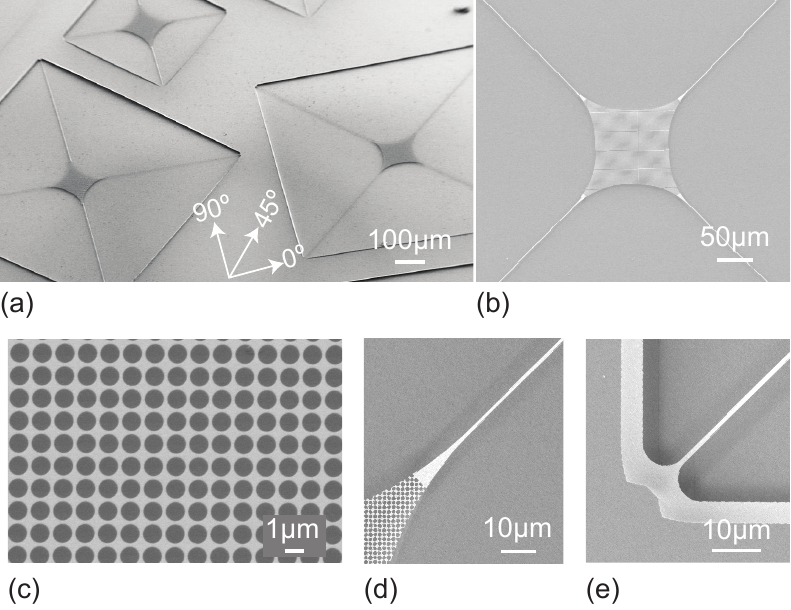}
    \caption{Tensile-strained trampoline resonators made from a \SI{73}{\nano\meter}-thick crystalline InGaP layer. Scanning electron microscope (SEM) images of (a) trampoline resonators oriented along different crystal directions and varied tether length (tether width \SI{1}{\micro\meter}, central pad area $100\, \times$ \SI{100}{\micro\meter}$^2$). The angles $0^{\circ}$, $45^{\circ}$, and $90^{\circ}$ denote the crystal directions $[1\,1\,0]$, $[1\,0\,0]$, and $[1\,\bar{1}\,0]$, respectively. (b) Close-up of a trampoline resonator with a tether length of \SI{750}{\micro\meter}. Enlarged views of the resonator in (b) showing (c) the photonic crystal (PhC) pattern on the central area with PhC hole radius $r_\textrm{PhC} =$ \SI{544}{\nano\meter} and period $a_\textrm{PhC} =$ \SI{1304}{\nano\meter}, (d) the tether connection to the central pad, and (e) the tether clamping to the substrate.}
    \label{fig:sem_1}
\end{figure}

The devices in this work were fabricated after stripping the top \InGaPx{} and $\textrm{Al}_{0.68}\textrm{Ga}_{0.32}\textrm{As}$ layers. From X-ray diffraction analysis we find that the gallium content in the as-grown InGaP layer is $0.53 \leq x \leq 0.59$, with a value of $x = 0.5658$ matching our inference of released stress from InGaP string resonators. Hence, we will assume throughout the rest of this work $x = 0.5658$ (and for abbreviation sometimes write $0.57$). We use electron-beam lithography to expose the resonator patterns to a resist stack containing an adhesion promoter and an electron beam resist. Chlorine-based RIE-ICP etching is then used to transfer the pattern onto the InGaP device layer followed by releasing the devices with a selective anisotropic wet etch using a mixture of citric acid and hydrogen peroxide \cite{Arslan1999}. Finally, we perform critical point drying to prevent the devices from collapsing due to capillary forces. \fref{fig:sem_1} shows fabricated trampoline resonators.

\section{Material properties}

Material properties of the InGaP layer, in particular its tensile stress, yield strength, and intrinsic mechanical quality factor, are key factors to engineer high-quality mechanical resonators at desired eigenfrequencies. We determine these material properties experimentally by fabricating string resonators, following methods introduced in Refs.~\cite{Buckle2018,Federov2019,bereyhiclamptapering2019,Buckle2021}.

\subsection{Tensile stress of InGaP string resonators}

The intrinsic stress in the \InGaPval{} layer originates from its lattice mismatch with the GaAs buffer layer. The in-plane as-grown strain of the thin \InGaPx{} layer is 
\begin{equation}\label{eq:latticestrain}
    \epsilon(x) = \frac{a_{\textrm{GaAs}} - a_{\textrm{In}_{1-x}\textrm{Ga}_{x}\textrm{P}}}{a_{\textrm{In}_{1-x}\textrm{Ga}_{x}\textrm{P}}},
\end{equation}
where $a_{\textrm{GaAs}}$ and $a_{\textrm{In}_{1-x}\textrm{Ga}_{x}\textrm{P}}$ are the lattice constants of GaAs and \InGaPx{}, respectively. One can tune the as-grown stress in the \InGaPx{} device layer by varying $x$ to obtain compressive stress for $x < 0.515$ and tensile stress for $x > 0.515$ (see \SM{}~\ref{app:materials}). The InGaP layer can be grown without defects until a certain critical thickness governed by $x$, which is \SI{1132}{\nano\meter} for $x = 0.5658$  \cite{People1985}. In our case, the InGaP device layer has an as-grown thickness of \SI{73}{\nano\meter}, which is well below this limit (see \SM{}~\ref{app:materials}). 

\begin{figure}[t!hbp]
    \centering\includegraphics{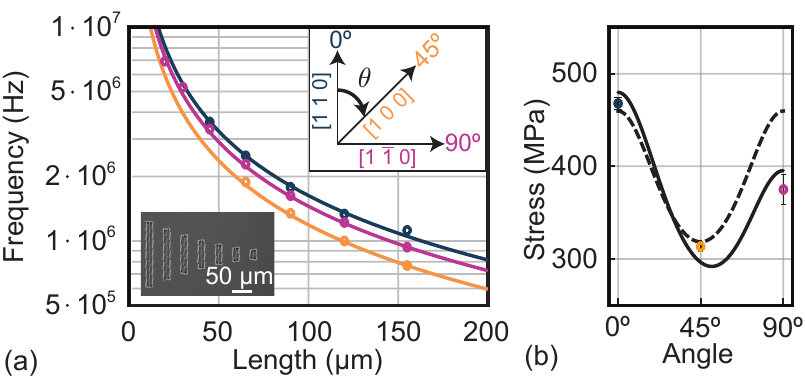}
    \caption{Tensile stress of the \SI{73}{\nano\meter}-thick \InGaPval{} string resonators. (a) Mechanical frequencies of the fundamental mode of string resonators of different lengths (with a width of \SI{200}{\nano\meter}) along the three crystal directions $[1\,1\,0]$, $[1\,0\,0]$, and $[1\,\bar{1}\,0]$ denoted as $0^{\circ}$, $45^{\circ}$, and $90^{\circ}$, respectively. The lines are a fit to the expected frequencies of tensile-strained string resonators (\eref{eq:stress_freq1}). The inset shows an optical microscope image of string resonators of different lengths oriented along  $0^{\circ}$. (b) The extracted tensile stress along different crystal directions is shown as points. The dashed line shows the tensile stress $\sigma(x,\theta)$ predicted from in-plane strain $\epsilon(x)$ and Young's modulus $E(x,\theta)$, see \eref{eq:sigma}. The solid line shows the tensile stress that takes into account an additional angle-dependent contribution to Young's modulus (see \eref{eq:angledepyoung}).}
    \label{fig:strings1}
\end{figure}

The crystalline structure of \InGaPx{} results in an orientation-dependent released axial stress $\sigma(x,\theta)$ \cite{Buckle2018} (for details see \SM{}~\ref{app:materials})
\begin{equation}\label{eq:sigma}
    \sigma(x,\theta) = E(x,\theta)\,\epsilon(x),
\end{equation}
where the angle $\theta$ is defined with respect to the crystal directions as shown in the inset of \fref{fig:strings1}. Importantly, $\sigma(x,\theta)$ is the released stress as Young's modulus $E(x,\theta)$ accounts for an anisotropic Poisson ratio (see \SM{}~\ref{app:materials}). We determine the anisotropic stress in \InGaPval{} from measurements of string resonators of different lengths oriented along different directions. The eigenmode frequencies of tensile-strained string resonators are given by \cite{schmid2016fundamentals}

\begin{equation}\label{eq:stress_freq1}
    f_n = \frac{n}{2L} \sqrt{\frac{\sigma(x,\theta)}{\rho(x)}},
\end{equation}
where $n$ is the mode number, $\rho(x)$ is the density of the material, and $L$ is the length of the resonator.

We fabricated string resonators with lengths between $\SI{20}{\micro\meter}$ and $\SI{160}{\micro\meter}$ and a width of $\SI{200}{\nano\meter}$ along the crystal directions  $[1\,1\,0]$, $[1\,0\,0]$, and $[1\,\bar{1}\,0]$. We measured their thermally-driven displacement noise power spectrum (NPS) in a high vacuum environment with an optical homodyne detection setup, for details see Ref.~\cite{Kini2020}. The same setup was used for characterizing the mechanical properties of the InGaP trampoline resonators. \fref{fig:strings1}(a) shows the measured fundamental eigenmode frequencies of the $\SI{200}{\nano\meter}$-wide string resonators. We use the measured eigenfrequencies of all identified eigenmodes to determine the stress along the different crystal directions using \eref{eq:stress_freq1} and $\rho(0.57)$. We obtain a released stress in the string resonators of $\sigma(0^{\circ}) = $\SI{467.7(71)}{\mega\pascal}, $\sigma(45^{\circ}) = $\SI{313.3(54)}{\mega\pascal} and $\sigma(90^{\circ}) =$ \SI{374.9(164)}{\mega\pascal}.

We can estimate the Ga content of the \InGaPx{} layer based on the experimentally determined stress values by using \eref{eq:sigma}. Note that this equation accounts for stress relaxation by incorporating Poisson's ratio in $E(x,\theta)$ (see \SM{}~\ref{app:Youngsmod}). We estimate a Ga content of $0.5667$, $0.5649$ and $0.5566$ from the stress along $0 ^{\circ}$, $45 ^{\circ}$, and $90 ^{\circ}$, respectively. As the Ga contents along $0^{\circ}$ and $45 ^{\circ}$ are similar, we use the average value of $x = 0.5658$ to estimate the expected crystal-direction dependent released stress, seen as the dashed-line in \fref{fig:strings1}(b). This prediction captures the released stress along $0 ^{\circ}$ and $45 ^{\circ}$, as expected, but not along $90^{\circ}$. However, from the crystal structure, we would expect the stress to be identical along the $0^{\circ}$ and $90^{\circ}$ directions. This unexpected deviation has also been observed in Ref.~\cite{Buckle2018}, which attributed it to a defect density that varies along different crystal directions. Alternatively, spontaneous ordering in MOCVD growth may also be a possible reason \cite{zakariaInfluenceDegreeOrder2010}. This modification of Young's modulus, $\Delta\,E(x,\theta)$, can be modeled with a $\cos(2\theta)$ function as
\begin{align}\label{eq:angledepyoung}
    \nonumber\Delta\,E(x,\theta) & = \sigma(\theta)/\epsilon(x) - E(x,\theta) \\
            & =  \alpha + \beta\cos(2\theta). 
\end{align}

We obtain $\alpha=-5.9\,\SI{}{\giga\pascal}$  and $\beta=11.3\,\SI{}{\giga\pascal}$ (in Ref.~\cite{Buckle2018} $\alpha=-5.5\,\SI{}{\giga\pascal}$, $\beta= 5.1\,\SI{}{\giga\pascal}$). The stress including the deviation $\Delta\,E(x,\theta)$ is shown as the solid line in \fref{fig:strings1}(b) and captures the data well. We attribute the remarkably small difference between our determined values for $\alpha$ and $\beta$ and the ones from Ref.~\cite{Buckle2018} to the difference in growth method (MOCVD vs.~MBE), the gallium content (0.5658 vs.~0.59), and the resonator's support geometry. Experimental determination of Young's modulus along the crystal directions of InGaP \cite{klassDeterminingYoungModulus2022} together with detailed material studies are required to explore the microscopic origin for this additional anisotropy.

\subsection{Yield strength of InGaP layer}

\begin{figure}[t!hbp]
\centering\includegraphics{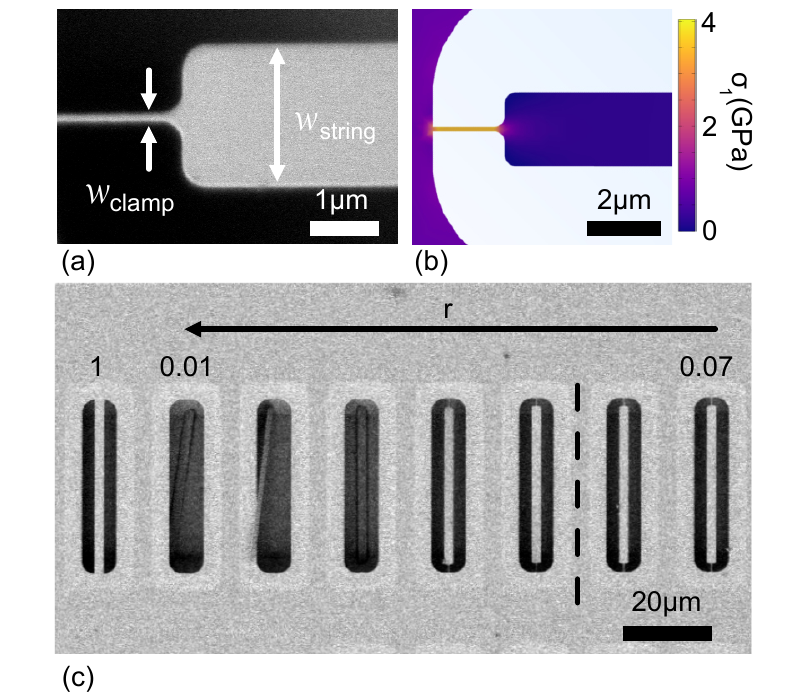}
    \caption{Determination of the yield strength of the \SI{73}{\nano\meter}-thick InGaP layer using tapered string resonators. (a) SEM image of the tapered clamp region. (b) FEM simulation of the first principal stress, $\sigma_1$, of the tapered string. (c) SEM image of tapered string resonators oriented along $0^\circ$. Resonators of $r<0.06$ (dashed line) fracture.}
    \label{fig:sem_yield_2}
\end{figure}

It is desirable to maximize the strain (respective stress) in the mechanical device layer to increase the effect of dissipation dilution. Since InGaP is a brittle material, a limit is set by the maximal applicable stress, i.e., $\sigma_{\mathrm{yield}}$($\theta$), after which the material breaks. We determine the yield strength of the InGaP layer experimentally following the method from Ref.~\cite{bereyhiclamptapering2019}. We fabricated string resonators that are attached to the substrate with a tapered clamp of varying ratio $r={w_{\textrm{clamp}}}/{w_{\textrm{string}}}$, where $w_{\textrm{clamp}}$ is the width of the string at the clamping point and $w_{\textrm{string}}$ is the width at the center of the string, see \fref{fig:sem_yield_2}(a). The width at the clamping point determines the strain redistribution along the string resonator. The strain in the clamping region can be obtained from the strain-width relation as $\epsilon_\textrm{clamp} = \epsilon_{r=1}\cdot w_{\textrm{string}}/w_{\textrm{clamp}} = \epsilon_{r=1}/r$ \cite{Federov2019}. As stress is proportional to strain (\eref{eq:sigma}), the stress at the clamping region is enhanced for $r<1$ [\fref{fig:sem_yield_2}(b)] and an increase above the yield stress of the material results in fracturing of the beam. The corresponding yield stress is (for details see Ref.~\cite{Federov2019})

\begin{equation}\label{eq:yieldstress}
    \sigma_\textrm{yield} = \sigma_{r=1}/r_\textrm{{yield}}.
\end{equation}

In our case, the yield stress depends on the orientation of the string resonators with respect to the crystal directions.

We fabricated 15 arrays of \SI{40}{\micro\meter}-long and \SI{2}{\micro\meter}-wide string resonators with $r$ varying from $0.001$ to $1$ along the three aforementioned crystal directions. We observe that strings oriented along $0^{\circ}$ [\fref{fig:sem_yield_2}(c)], $45^{\circ}$, and $90^{\circ}$ break for $r_\textrm{{yield}}$ of \SI{0.062(5)}, \SI{0.061(7)}, and \SI{0.067(5)}, respectively. We determined $\sigma_{r=1}$ from a measurement of the fundamental mode frequency of the $r=1$ resonator and using \eref{eq:stress_freq1}. We obtain yield stresses $\sigma_\textrm{{yield}}$ using \eref{eq:yieldstress} of \SI{5.5(8)}{\giga\pascal}, \SI{3.3(5)}{\giga\pascal}, and \SI{3.7(5)}{\giga\pascal} along $0^{\circ}$, $45^{\circ}$, and $90^{\circ}$, respectively. The corresponding yield strain of $\epsilon_\textrm{yield}(\theta) = \sigma_\textrm{{yield}}(\theta)/[E(\theta)+\Delta E(\theta)]$ is \SI{0.043(8)}{}, \SI{0.041(8)}{}, and \SI{0.034(5)}{}, along $0^{\circ}$, $45^{\circ}$, and $90^{\circ}$, respectively. The obtained yield strength is comparable to Si$_{3}$N$_{4}$ (\SI{6}{\giga\pascal}), but lower than the one of SiC (\SI{21}{\giga\pascal}) or Diamond (\SI{35}{\giga\pascal}) \cite{sementilli_nanomechanical_2022}.

To verify the analytic model for determining the yield strength, we perform independent FEM simulations of the corresponding tapered geometries [\fref{fig:sem_yield_2}(b)] (parameters see \SM{}~\ref{app:params}). Using $r_\textrm{{yield}}$, FEM simulations predict yield stresses of $\sigma_\textrm{{yield}}(0^{\circ}, 90^{\circ}) \approx$ \SI{4.14}{\giga\pascal}, $\sigma_\textrm{{yield}}(45^{\circ}) \approx$ \SI{3.2}{\giga\pascal} and yield strain of $\approx \SI{0.029}{}$, in reasonable agreement with the experimental values.

\subsection{Intrinsic quality factor from InGaP string resonators}
\label{sec:intrinsicQ}

The quality factor $Q$ of a mechanical resonator is generally given by
 \cite{imbodenDissipationNanoelectromechanicalSystems2014}
\begin{equation}
    Q^{-1} =  Q_{\mathrm{int}}^{-1} +Q_{\mathrm{ext}}^{-1},
\end{equation}
where $Q_{\mathrm{int}}$ and $Q_{\mathrm{ext}}$ are the quality factors limited by intrinsic and extrinsic loss mechanisms, respectively. In the following, we determine $Q_{\mathrm{int}}$, which captures material-related loss processes of the \SI{73}{\nano\meter}-thick InGaP layer. To this end, we use string resonators and we confirmed that they are not limited by clamping loss or gas damping  (\SM{}~\ref{app:damping}), which determine $Q_{\mathrm{ext}}$ in our case.

\subsubsection{Dissipation dilution}

We determine $Q_{\mathrm{int}}$ of the \SI{73}{\nano\meter} InGaP layer from measurements of the quality factor of strained InGaP string resonators. Importantly, the stress in the string resonators dilutes $Q_{\mathrm{int}}$ by a factor $D$ \cite{schmid2016fundamentals,Federov2019,sementilli_nanomechanical_2022} 
\begin{equation}\label{eq:Qd}
    Q_{\textrm{D}} = D \cdot  Q_{\textrm{int}}.    
\end{equation}

The dilution factor $D$ depends on the stress, resonator geometry, and displacement mode profile. For a uniform string resonator, one obtains \cite{Yu2012,Federov2019}
\begin{equation}\label{eq:Dil_fac_strings}
    D_n = \frac{1}{2 \lambda + (\pi n)^2 \lambda^2},
\end{equation}
where $n$ is the mode number and $\lambda$ is a stress parameter given as 
\begin{equation}
    \lambda = \frac{h}{L} \sqrt{\frac{E}{12\, \sigma}},
\end{equation}
with length $L$ and thickness $h$ of the string resonator. 

\begin{figure}[t!hbp]
    \centering\includegraphics{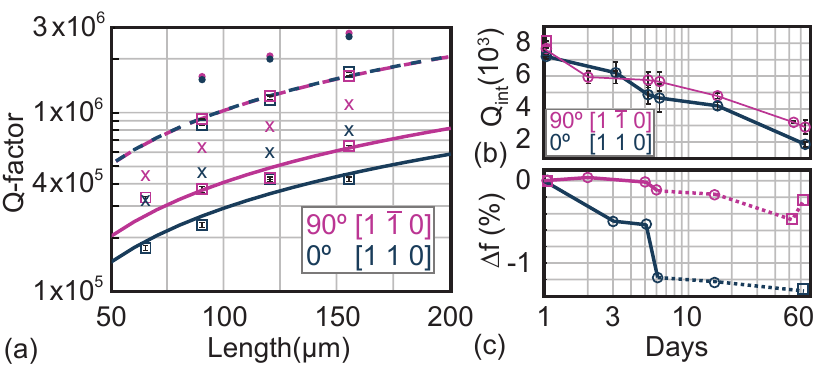}
    \caption{Determination of the intrinsic mechanical quality factor using InGaP string resonators. (a) Measured $Q$ factors (squares) for the fundamental mode of string resonators with varying lengths oriented along $0^{\circ}$ and $90^{\circ}$. The dashed (solid) lines are fits to extract $Q_{\mathrm{int}}$ at day 1 (day 60). The dots (crosses) show $Q_{\textrm{D}}^{\mathrm{FEM}}$ obtained from FEM, when using $Q_{\mathrm{int}}$ from day 1 (day 60) as input. (b) We observe that $Q_{\mathrm{int}}$ degrades over time, shown for two samples (squares and circles). The solid (dashed) line indicates when the sample was stored in vacuum (ambient condition). (c) The relative change of the resonance frequency of the string resonators, i.e., $\Delta f=(f_{\mathrm{day\,x}}-f_{\mathrm{day\,1}})/f_{\mathrm{day\,1}}$, is shown over the same time period.}
    \label{fig:Qint}
\end{figure}

We fabricated strings of different lengths with a width of \SI{2}{\micro\meter} oriented along different crystal directions to infer $Q_{\textrm{int}}$. \fref{fig:Qint}(a) shows the measured quality factors for the fundamental mode extracted from ringdown measurements. Using $D_1$, we obtain $Q_{\textrm{int}}$ of $7550 \pm 140$ and $8150 \pm 320$ along $0^{\circ}$ and $90^{\circ}$, respectively. Our determined average $Q_{\textrm{int}}$ of about $7.9\cdot 10^3$ for the \SI{73}{\nano\meter}-thick InGaP layer is comparable to LPCVD-grown Si$_3$N$_4$ (\SI{66}{\nano\meter}-thick: $3.75\cdot 10^3$ \cite{villanueva2014evidence}, with $6.9\cdot 10^3\cdot h \textrm{[100\,\textrm{nm}]}$ \cite{villanueva2014evidence,Federov2019} for \SI{73}{\nano\meter} SiN: $Q_{\textrm{int}} = 5\cdot 10^3$) and \SI{14}{\nano\meter}-thick s-Si ($8\cdot 10^3$ \cite{beccari2022strained}), and larger than for \SI{75}{\nano\meter}-thick SiC ($1.5\cdot 10^2$ \cite{romero2020engineering}).

\subsubsection{Dilution factor in FEM}

The dilution factor can be numerically computed using FEM simulations (see \SM{}~\ref{app:damping}) \cite{Federov2019, Fedorov_Thesis,bereyhi2022hierarchical}. This approach is required when analyzing dissipation dilution of more complex mechanical resonator geometries, as in our case, trampoline mechanical resonators including a PhC pattern. We verify the FEM approach by simulating the dilution factor $D_{\mathrm{FEM}}$ for string resonators. Using the experimentally determined intrinsic quality factor, $Q_{\mathrm{int}}$, we calculate $Q_{\textrm{D}}^{\mathrm{FEM}}=D_{\mathrm{FEM}}Q_{\mathrm{int}}$, which are shown in \fref{fig:Qint}(a). We find that $Q_{\textrm{D}}^{\mathrm{FEM}}$ is slightly larger than the measured $Q$ factors, similar to other works \cite{romero2020engineering,bereyhi2022hierarchical,Bereyhi2022Perimeter}.

\subsubsection{Degradation of the intrinsic quality factor with time}

We observe that the mechanical quality factor of the string resonators decreases with time. \fref{fig:Qint}(b) shows this change of $Q_{\textrm{int}}$ for two different samples. The samples were measured in high vacuum, but they were stored in between measurements either in vacuum or under ambient conditions, see \fref{fig:Qint}(b). $Q_{\textrm{int}}$ decreases over two months by up to a factor of 4. During the same time period the resonance frequency changes by less than $2\%$, see \fref{fig:Qint}(c). We hypothesize that the degradation of the quality factor is not due to a gradual relaxation of the tensile stress of the InGaP layer. At the moment, we can only speculate that the InGaP layer undergoes some modification, for example, moisture-induced degradation \cite{kimUltraLightweightFlexibleInGaP2021} or other processes \cite{bahlReliabilityInvestigationInGaP1995} that may lead to an increase of mechanical dissipation. Future work is required to determine the cause of the InGaP degradation. To this end, the InGaP layer can be examined periodically with X-ray diffraction (to obtain information about the formation of an oxide surface layer) and photoemission spectroscopy (to obtain data on the elements present on the surface and their chemical state). Scanning near-field optical microscopy, micro-Raman, or tip-enhanced Raman spectroscopy can be used on the InGaP nanomechanical resonators to obtain, e.g., spatial information about potential strain changes. Mitigation strategies include surface passivation \cite{gorbylevHydrogenPassivationEffects1994} or capping of the InGaP layer with thin GaAs layers.

\section{Trampoline resonators in InGaP}

For efficient transduction of mechanical motion to out-of-plane light, the reflectivity of the mechanical resonator is desired to be close to unity. At the same time, the thickness of the device layer should be sufficiently thin to keep mechanical damping small. These requirements can be fulfilled by patterning thin mechanical resonators with a PhC \cite{buiHighreflectivityHighQMicromechanical2012,makles2DPhotoniccrystalOptomechanical2015,norte2016mechanical,bernardPrecisionResonanceTuning2016,gartner2018integrated,Kini2020}. We, therefore, choose a trampoline-shaped geometry, which allows patterning its central area with a PhC to achieve the desired reflectivity and at the same time allows decreasing mechanical dissipation by use of dissipation dilution \cite{norte2016mechanical,reinhardtUltralowNoiseSiNTrampoline2016}, see \fref{fig:trampolineNPS}(b).

\subsection{Mechanical properties of InGaP trampoline resonators}

Refs.~\cite{norte2016mechanical,reinhardtUltralowNoiseSiNTrampoline2016} demonstrated that high-$Q$ trampolines can be realized with thin and long tethers that connect the central pad to the support. In our work, we can reliably fabricate InGaP trampolines with a tether width of \SI{1}{\micro\meter} and tether length of up to \SI{750}{\micro\meter} [\fref{fig:sem_1}(b)], and with a radius of \SI{10}{\micro\meter} at the tether clamp to the support [\fref{fig:sem_1}(e)] and of \SI{200}{\micro\meter} at the tether clamp to the pad [\fref{fig:sem_1}(d)].

\begin{figure}[t!hbp]
    \centering\includegraphics{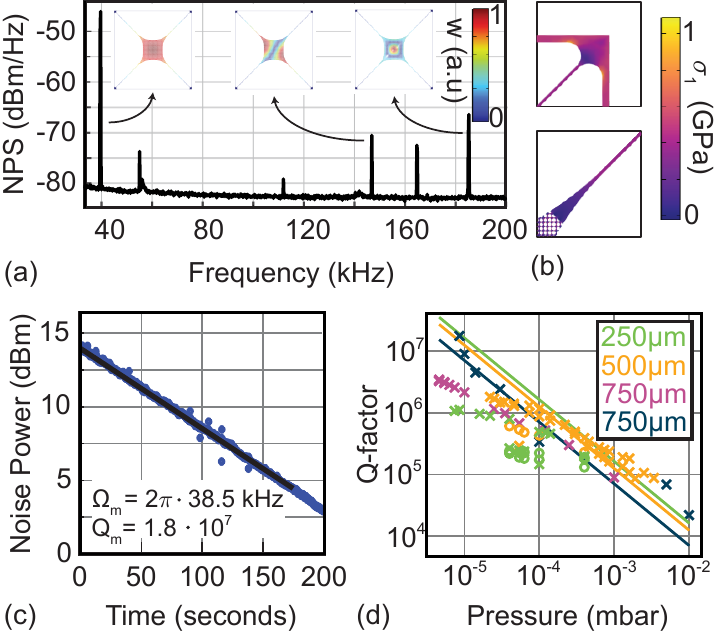}
    \caption{Mechanical properties of InGaP trampoline resonators. (a) Noise power spectrum (NPS) of a trampoline resonator of \SI{750}{\micro\meter} tether length, \SI{1}{\micro\meter} tether width and central PhC pad size of $100\, \times$ \SI{100}{\micro\meter}$^2$. The insets show FEM simulated mode shapes depicting the out-of-plane displacement $w$. (b) FEM simulations of the first principal stress in the released device at the tether connection to the pad and the clamping region. (c) Ringdown measurement on day five of the fundamental mode of the trampoline from (a), which was placed in vacuum directly after fabrication on day one. We obtain a $Q$ of $1.8\cdot10^7$ from a fit (solid line) to the decay. (d) The dependence of $Q$ on pressure for the fundamental mode of trampolines of different lengths with tethers oriented along different crystal directions (0$^\circ$/90$^\circ$ marked by a circle, 45$^\circ$/135$^\circ$ by a cross). The solid lines show the quality factor limited by gas damping for trampolines of tether length \SI{250}{\micro\meter}, \SI{500}{\micro\meter}, and \SI{750}{\micro\meter}.}
    \label{fig:trampolineNPS}
\end{figure}

\fref{fig:trampolineNPS}(a) shows a thermally-driven displacement noise power spectrum of an InGaP trampoline resonator. The tethers of this device are \SI{750}{\micro\meter}-long and oriented along $45^{\circ}/135^{\circ}$. We observe the fundamental mode at \SI{38.5}{\kilo\hertz} and several higher-order modes, which we identify by comparing measured eigenfrequencies to the ones simulated via FEM. \tref{tab:FEMtrampoline} shows measured fundamental mode eigenfrequencies for trampolines with various tether lengths. As expected, we find that trampolines with shorter tether lengths exhibit higher resonance frequencies. We observe that trampolines whose tethers are oriented along $0^{\circ}/90^{\circ}$ have larger frequencies than the ones oriented along $45^{\circ}/135^{\circ}$. We can understand this behaviour as the stress along $0^{\circ}/90^{\circ}$ is larger than the one along $45^{\circ}/135^{\circ}$ [see \fref{fig:strings1}(b)] resulting in a higher resonance frequency. We find a good agreement to the eigenfrequencies calculated with FEM. In the FEM simulations, we take into account the anisotropy of Young's modulus (\eref{eq:sigma}), but do not account for its deviation (\eref{eq:angledepyoung}), which is a possible reason for the small discrepancy between the FEM and measurement results.

The highest mechanical $Q$ factor that we measure is $1.8\cdot 10^{7}$ for the fundamental mode of the InGaP trampoline with \SI{750}{\micro\meter} tether length [see \fref{fig:trampolineNPS}(c)], resulting in a $Q\cdot f$ product of $7 \cdot 10^{11}\,$Hz. We measured this value at room temperature at a pressure of $8 \cdot 10^{-6}$\,mbar, which is close to the minimal achievable pressure that we can reach in our setup. With the current devices, we reach a calculated thermal noise limited force sensitivity of 50\,aN/$\sqrt{\mathrm{Hz}}$. When compared to SiN-based membrane-type devices at room temperature, our value lies in the same order of magnitude as reached with phononic band gap SiN membranes (37\,aN/$\sqrt{\mathrm{Hz}}$ \cite{halgMembraneBasedScanningForce2021}) and SiN trampolines (19.5\,aN/$\sqrt{\mathrm{Hz}}$ \cite{norte2016mechanical,reinhardtUltralowNoiseSiNTrampoline2016}).

We already noticed that the mechanical $Q$ of the InGaP string resonators decreases with time. We observe the same trend for the InGaP trampoline resonators. This behavior complicates a definite identification of the loss mechanism that limits the $Q$ of InGaP trampolines. Nevertheless, we look at different mechanical damping mechanisms in the following to analyze limits in achieving even higher $Q$. We consider gas damping first. To this end, we performed pressure-dependent measurements of various trampolines; the results are shown in \fref{fig:trampolineNPS}(d). We observe a linear increase of $Q$ with a decrease in pressure, as expected from gas damping \cite{schmid2016fundamentals} (\SM{}~\ref{app:damping}). The $Q$-factor of the \SI{750}{\micro\meter}-long trampoline (dark blue crosses) that was fabricated and immediately measured follows this gas-damping prediction. However, for samples that were measured with a delay after fabrication [other colors in \fref{fig:trampolineNPS}(d)], we observe a deviation from the gas damping limit at lower pressures, which indicates that the $Q$ factor of these trampolines reaches another limiting mechanism. As reaching low pressures requires some days of pumping, this deviation may originate from the degradation of $Q_{\mathrm{int}}$ over time. The amount of dissipation dilution achieved with the trampoline geometry may also limit the maximally achievable $Q$. To evaluate this, we computed $D_Q$ via FEM and obtain a value of $D_Q=1750$ for the \SI{750}{\micro\meter} tethered trampoline. With $Q_{\mathrm{int}}=7.9\cdot 10^3$ we obtain $Q_D\sim1.38\cdot10^7$. This $Q$ factor is close to the experimentally obtained result. Hence, the trampoline resonators may currently be limited by the achievable gas pressure or by the amount of dissipation dilution. Stabilization of $Q_{\textrm{int}}$ is required to identify with certainty the limiting damping mechanism and apply strategies to further reduce it.

\begin{table}[h!tbp]
\begin{tabular}{cccc}
    \hline\hline
    \multicolumn{2}{c}{Tether}                             & \multicolumn{2}{c}{Frequency (\SI{}{\kilo\hertz})}            \\ \hline
    \multicolumn{1}{c}{Length (\SI{}{\micro\meter})}               & Orientation & \multicolumn{1}{c}{Measured} &  Simulated\\ \hline
    \multicolumn{1}{c}{\multirow{2}{*}{250}} & $0^{\circ}/90^{\circ}$           & \multicolumn{1}{c}{90.9 }         & 106.9       \\ \cline{2-4} 
    \multicolumn{1}{c}{}                     & $45^{\circ}/135^{\circ}$          & \multicolumn{1}{c}{80}         &    90.5     \\ \hline
    \multicolumn{1}{c}{\multirow{2}{*}{500}} & $0^{\circ}/90^{\circ}$           & \multicolumn{1}{c}{54.2 }         &  62.3      \\ \cline{2-4} 
    \multicolumn{1}{c}{}                     & $45^{\circ}/135^{\circ}$          & \multicolumn{1}{c}{43.7 }         &   52.4     \\ \hline
    \multicolumn{1}{c}{\multirow{2}{*}{750}} & $0^{\circ}/90^{\circ}$           & \multicolumn{1}{c}{40.1 }         &  47.8      \\ \cline{2-4} 
    \multicolumn{1}{c}{}                     & $45^{\circ}/135^{\circ}$          & \multicolumn{1}{c}{38.5}         &     40.1    \\ \hline\hline
\end{tabular}
\caption{Measured and FEM-simulated eigenfrequencies of the fundamental mode of trampolines with varied tether length and orientation.}
\label{tab:FEMtrampoline}
\end{table}

\subsection{Optical reflectance}

\begin{figure}[t!hbp]
    \centering\includegraphics{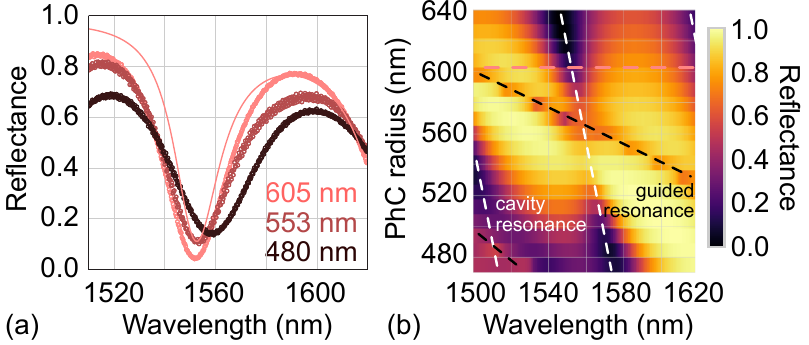}
    \caption{Reflectance spectra of InGaP trampoline resonators patterned with a PhC. (a) Measurements for $r_\textrm{PhC}$ of \SI{480}{\nano\meter}, \SI{553}{\nano\meter}, and \SI{605}{\nano\meter} (dots) and RCWA simulation for $r_\textrm{PhC}= \SI{605}{\nano\meter}$ (solid line). (b) Simulation of a reflectance map when varying the PhC radius. Other parameters are $a_\textrm{PhC} = $ \SI{1309}{\nano\meter}, $d_\textrm{PhC} = $ \SI{73}{\nano\meter}, vacuum gap \SI{14.8}{\micro\meter}, and the GaAs substrate is a semi-infinite layer.}
    \label{fig:trampoline_ref}
\end{figure}

In the following, we characterize the optical reflectance of the trampoline resonators patterned with a PhC. For details of the measurement setup, we refer the reader to Ref.~\cite{Kini2020}. \fref{fig:trampoline_ref}(a) shows reflectance spectra of three trampolines with square PhC patterns of lattice constant $a_\textrm{PhC} = \SI{1309}{\nano\meter}$ and PhC radii $r_\textrm{PhC}$ of \SI{480}{\nano\meter}, \SI{553}{\nano\meter}, and \SI{605}{\nano\meter}. We evaluated the PhC parameters via image recognition applied to high-magnification SEM images of the respective PhC pattern after fabrication. We observe that the PhC trampolines demonstrate an engineered reflectance in the wavelength range of $1510-1620$\SI{}{\nano\meter} with a pronounced modulation. The latter can be understood by noting that the trampoline is separated from the GaAs substrate by a vacuum gap of about \SI{15}{\micro\meter}, originating from the release of the trampoline in the wet etch fabrication step. This gap forms a low-quality optical cavity between the trampoline and the GaAs substrate.

This interpretation is supported by rigorous coupled wave analysis (RCWA) simulations of our system \cite{liu2012s4,Kini2020} (for parameters see \SM{}~\ref{app:params}). \fref{fig:trampoline_ref}(b) shows a simulated reflectance map when varying the PhC radius. We observe pronounced dips in reflectivity when hitting the cavity resonance condition. The simulated free spectral range is about \SI{75}{\nano\meter}, which is close to the expected value given by the gap and noting that the PhC additionally modifies the effective cavity length \cite{pottier2012evolution}. When decreasing $r_{\mathrm{PhC}}$, the cavity dip shifts to longer wavelengths implying an increased cavity length. This effect is also seen in the measurements, \fref{fig:trampoline_ref}(a). Another dip occurs in the reflectance map, which originates from the coupling of focused light into a guided resonance of the PhC \cite{mouraCentimeterscaleSuspendedPhotonic2018,Kini2020}.

We observe some discrepancies between measurement and simulation results, which we trace back to simplifying assumptions that we make in the RCWA simulations. The simulations assume an infinite and uniform PhC pattern. However, in the experiment, the PhC pattern is finite and exhibits unavoidable fabrication-related non-uniformity. These non-idealities lead to a reduction of the overall reflectivity of the PhC trampoline \cite{Kini2020}. Furthermore, the simulation assumes a fixed gap between PhC and substrate, which is not the case for the fabricated samples. We observe instead that the etched substrate surface exhibits an elevated and spatially varying height profile, which resembles the shape of the PhC trampoline (see \SM{}~\ref{app:materials} for an SEM image). As a result, we expect the cavity dip to be spectrally broadened due to a spatially varying gap length and due to scattering loss from the uneven substrate surface. The latter will also reduce the reflectivity of the combined system. To mitigate the reflectance features introduced by the gap, one can back-etch the sample to obtain free-standing PhC trampolines \cite{norte2016mechanical,reinhardtUltralowNoiseSiNTrampoline2016}. Alternatively, one can integrate an etch stop-layer \cite{Kini2020} or a high-reflectivity distributed Bragg reflector below the PhC trampoline to realize an integrated cavity optomechanical system.

\section{Conclusion and Outlook}
We have demonstrated that trampoline-shaped micromechanical resonators in tensile-strained \SI{73}{\nano\meter}-thick InGaP exhibit mechanical quality factors surpassing $10^7$ at room temperature at pressures of $8\cdot10^{-6}$\,mbar, resulting in a $Q\cdot f$ product of $7\cdot 10^{11}$\,Hz. An enhancement by a factor of $10$ would place the presented InGaP trampoline mechanical resonators in the regime of quantum optomechanics at room temperature \cite{norte2016mechanical}. The trampoline resonator was patterned with a PhC to engineer its out-of-plane reflectivity. We observed that the intrinsic mechanical quality factor of the InGaP mechanical resonators decreased over time. This undesired effect should receive future attention and may require surface passivation techniques \cite{gorbylevHydrogenPassivationEffects1994}. Once this issue is solved, mechanical dissipation in InGaP resonators can be further reduced by a simple increase of the tether length of the trampoline \cite{norte2016mechanical,reinhardtUltralowNoiseSiNTrampoline2016}, or by applying more sophisticated methods such as hierarchical clamping structures \cite{bereyhi2022hierarchical}, machine-learning supported engineering of mechanical dissipation \cite{hojUltracoherentNanomechanicalResonators2021,shin2022spiderweb}, quasi-phononic band gaps \cite{tsaturyan2017ultracoherent}, or density phononic crystal engineering \cite{hojUltracoherentNanomechanicalResonators2022}. Notably, the InGaP mechanical device layers can be incorporated in (Al,Ga)As heterostructures via epitaxial layer growth. This approach would allow the realization of integrated free-space cavity optomechanical systems in a crystalline material platform (see \SM{}~\ref{app:appsfuture}). Such compact optomechanical systems could implement bound-states in the continuum-based optomechanics \cite{fitzgeraldCavityOptomechanicsPhotonic2021}, multielement \cite{xuerebStrongCouplingLongRange2012} or hybrid optomechanical systems \cite{midoloNanooptoelectromechanicalSystems2018} on a chip. 

The data used in this work can be found in the open-access Zenodo database: https://doi.org/10.5281/zenodo.7441332 \cite{zenododata}.

\begin{acknowledgments}
We gratefully acknowledge Eva Weig, Nils Johan Engelsen, and Claus G\"artner for insightful discussions, Max Trippel and Tommy M{\"u}ller for support in sample growth, and Joachim Ciers for ellipsometer measurement. This work was supported in part by the \mbox{QuantERA} project C’MON-QSENS!, the Knut and Alice Wallenberg Foundation through a Wallenberg Academy Fellowship (W.W.), by the Wallenberg Center for Quantum Technology (WACQT, A.C.), by Chalmers Excellence Initiative Nano, and by the Swedish Research Council (Grant no.~2019-04946). Sample fabrication was performed in the Myfab Nanofabrication Laboratory at Chalmers. Simulations were performed on resources provided by the Swedish National Infrastructure for Computing (SNIC) at Tetralith, Link{\"o}ping University, partially funded by the Swedish Research Council (Grant 2018-05973).
\end{acknowledgments}

\clearpage

\appendix

\section{Fabrication and InGaP material properties}
\label{app:materials}

\subsection{Fabrication}

\begin{figure}[t!hbp]
    \centering\includegraphics{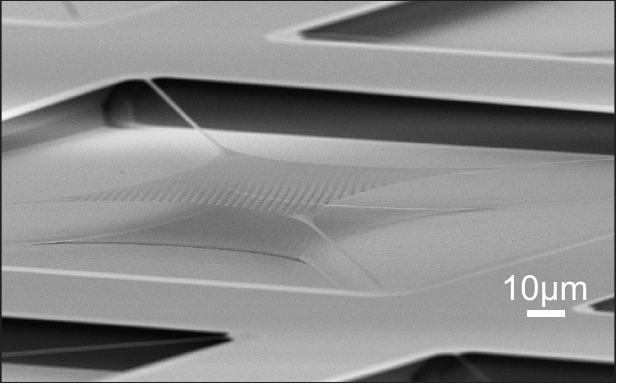}
    \caption{Tilted SEM image of a PhC trampoline resonator. The etched GaAs substrate under the trampoline is clearly seen to exhibit an elevated and uneven surface with a height profile mimicking the PhC trampoline.}
    \label{fig:trampolineshadow}
\end{figure}

\fref{fig:trampolineshadow} shows an SEM image of a fabricated and suspended InGaP trampoline. One can clearly see that the etched GaAs substrate surface under the trampoline is elevated and uneven. The unevenness will contribute to scattering loss of the optical cavity mode that forms between the suspended PhC trampoline and the substrate surface. The reason for this unevenness comes from the wet etch process. The PhC pattern determines the height profile of the "imprinted" surface onto the substrate. Basically, the wet etch process is faster on the large open areas outside of the trampoline pad than through the small diameter PhC holes. As a result, the material below the trampoline is initially etched slower than the material outside of the pad. At the end of the wet etch process, one obtains an elevated height profile mimicking the shape of the trampoline on the substrate surface.

\subsection{Critical thickness}

\begin{figure}[t!hbp]
    \centering\includegraphics{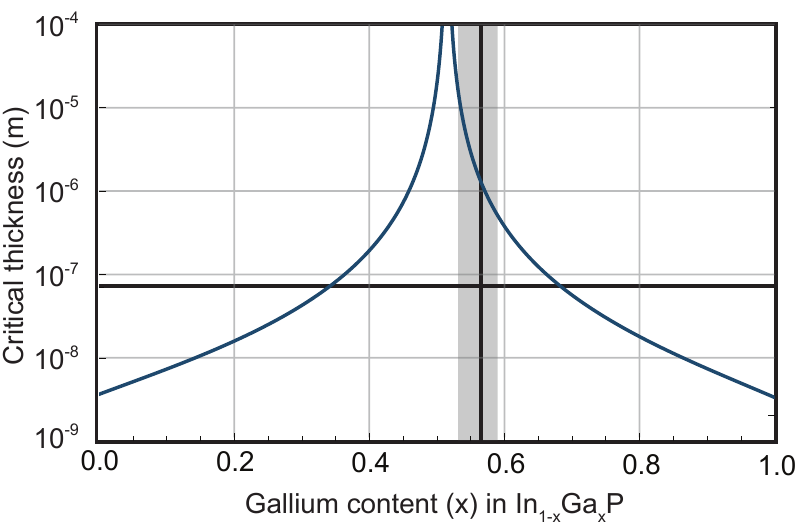}
    \caption{Critical thickness of an \InGaPx{} layer grown on GaAs calculated using the People and Bean model \cite{People1985}. The horizontal line represents the thickness of InGaP in this work, (\SI{73}{\nano\meter}). The gray area represents the range of gallium content estimated via X-ray diffraction and the vertical line represents the content that we use in this work ($x = 0.5658$).}
    \label{fig:CriticalThickness}
\end{figure}

\fref{fig:CriticalThickness} shows the critical thickness of \InGaPx{} on GaAs using the People and Bean model \cite{People1985}. We observe that the thickness of \InGaPval{} from our work is well below the critical thickness criterion for the corresponding gallium content and hence the device layer does not relax to its native lattice constant and will therefore be tensile strained.

\subsection{Anisotropic elasticity}
\label{app:Youngsmod}

The general linear relation between the components of the stress tensor $\sigma_{ij}$ and strain tensor $\epsilon_{kl}$ is given by the fourth order stiffness tensor $C_{ijkl}$ as
\begin{equation}
    \sigma_{ij} = C_{ijkl}\epsilon_{kl}.
\end{equation}

As InGaP has a zincblende crystal structure with a cubic symmetry, one can relate stress and strain with a reduced number of independent components by
\begin{equation}
    \begin{pmatrix}
    \sigma_{xx}\\
    \sigma_{yy}\\
    \sigma_{zz}\\
    \sigma_{yz}\\
    \sigma_{xz}\\
    \sigma_{xy}\\
    \end{pmatrix}
     = 
    \begin{pmatrix}
    c_{11} & c_{12} & c_{12} & 0 & 0 & 0\\
    c_{12} & c_{11} & c_{12}& 0 & 0 & 0\\
    c_{12} & c_{12} & c_{11}& 0 & 0 & 0\\
    0 & 0 & 0 & c_{44} & 0 & 0\\
    0 & 0 & 0 & 0 & c_{44} & 0\\
    0 & 0 & 0 & 0 & 0 & c_{44}\\
    \end{pmatrix}
    \begin{pmatrix}
    \epsilon_{xx}\\
    \epsilon_{yy}\\
    \epsilon_{zz}\\
    \epsilon_{yz}\\
    \epsilon_{xz}\\
    \epsilon_{xy}\\
    \end{pmatrix},
    \label{mat:C_cub}
\end{equation}
where $x,y,z$ denote the crystal directions [$1\,0\,0$], [$0\,1\,0$], and [$0\,0\,1$], respectively.
The elastic constants of \InGaPx{}, $c_{11}(x), c_{12}(x), c_{44}(x)$ \cite{shur1996handbook}, are listed in \tref{tab:params}. 
The simplified matrix for $C$ from \eref{mat:C_cub} can be expressed in terms of material constants ($\nu_{ij} = -\epsilon_{jj}/\epsilon_{ii}, E_i=\sigma_{ii}/\epsilon_{ii}, G_{ij}=\sigma_{ij}/\epsilon_{ij}$) \cite{hopcroft19young} as
\begin{equation}
    C
     = 
    \begin{pmatrix}
    \frac{1-\nu_{yz}\nu_{zy}}{E_yE_z\Delta} & \frac{\nu_{yx} + \nu_{yz}\nu_{zy}}{E_yE_z\Delta} & \frac{\nu_{zx} + \nu_{yz}\nu_{zy}}{E_yE_z\Delta} & 0 & 0 & 0\\
    \frac{\nu_{xy}+\nu_{xz}\nu_{zy}}{E_yE_x\Delta} & \frac{1 - \nu_{yx}\nu_{xz}}{E_yE_x\Delta} & \frac{\nu_{zy} + \nu_{yx}\nu_{xy}}{E_xE_z\Delta} & 0 & 0 & 0\\
    \frac{\nu_{xz}+\nu_{xy}\nu_{yz}}{E_xE_y\Delta} & \frac{\nu_{yz} + \nu_{xz}\nu_{yx}}{E_xE_y\Delta} &  \frac{1 - \nu_{xy}\nu_{yx}}{E_xE_y\Delta} & 0 & 0 & 0\\
    0 & 0 & 0 & G_{yz} & 0 & 0\\
    0 & 0 & 0 & 0 & G_{zx} & 0\\
    0 & 0 & 0 & 0 & 0 & G_{xy}\\
    \end{pmatrix}
\end{equation}
where $\Delta = \frac{1 - \nu_{xy}\nu_{yx} - \nu_{yz}\nu_{zy} - \nu_{zx}\nu_{zx} - 2\nu_{xy}\nu_{yz}\nu_{xz}}{E_x E_y E_z}$.

The inverse of the stiffness matrix is the compliance matrix and relates strain to stress, $\epsilon_{ij}=S_{ijkl}\sigma_{kl}$, via
\begin{equation}
    S =
    \begin{pmatrix}
    \frac{1}{E_x} & -\frac{\nu_{yx}}{E_y} & -\frac{\nu_{zx}}{E_z} & 0 & 0 & 0\\
    -\frac{\nu_{xy}}{E_x} & \frac{1}{E_y} & -\frac{\nu_{zy}}{E_z} & 0 & 0 & 0\\
    -\frac{\nu_{xz}}{E_x} & -\frac{\nu_{yz}}{E_y} &  \frac{1}{E_z} & 0 & 0 & 0\\
    0 & 0 & 0 & \frac{1}{G_{yz}} & 0 & 0\\
    0 & 0 & 0 & 0 & \frac{1}{G_{zx}} & 0\\
    0 & 0 & 0 & 0 & 0 & \frac{1}{G_{xy}}\\
    \end{pmatrix}
\end{equation}

\begin{figure}[b!htp]
    \centering\includegraphics{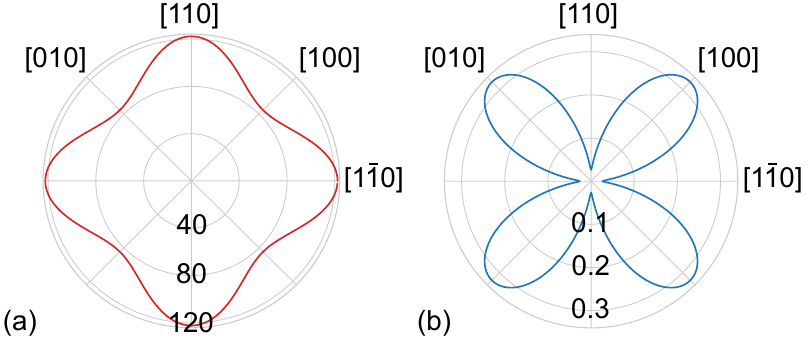}
    \caption{Theoretical values for (a) Young's modulus (in GPa) and (b) Poisson's ratio.}
    \label{fig:E_nu}
\end{figure}

Thus, one can easily obtain Young's modulus in the $x$ direction as $E_x = 1/s_{11}$. By rotating the matrices around the [$0\,0\,1$]-direction \cite{wortman1965young}, Young's modulus in the ($0\,0\,1$) plane can be obtained as \cite{Buckle2018}
\begin{widetext}
    \begin{multline}
        E(x,\theta) = \\
        = \frac{8\left[c_{11}(x)-c_{12}(x)\right]\left[c_{11}(x)+2c_{12}(x)\right]c_{44}(x)}{c_{11}^2(x)-2c_{12}(x)\left[c_{12}(x)-2c_{44}(x)\right]+c_{11}(x)\left[c_{12}(x)+6c_{44}(x)\right]+\left[c_{11}(x)+2c_{12}(x)\right]\left[c_{11}(x)-c_{12}(x)-2c_{44}(x)\right]\cos{(4\theta)}}.
    \end{multline}
\end{widetext}
For a gallium content of $x = 0.5658$, $E(x,\theta)$ is plotted in Fig.~\ref{fig:E_nu}(a).

As we are interested in the effect of the stress component along the string, we consider Poisson's ratio $\nu=\nu_{xy}$. This value can be obtained in the ($0\,0\,1$) plane from the transformed compliance matrix as $\nu = -s_{21}/s_{11}$. The resulting Poisson's ratio is shown in \fref{fig:E_nu}(b).

We assume that the as-grown strain along 0$^\circ$ and 90$^\circ$ is the same, i.e., $\epsilon_{xx} = \epsilon_{yy} = \epsilon$. Then, the as-grown stress of \InGaPx{} for $x = 0.5658$ is $\sigma_{\mathrm{grown}} = \sigma_{xx} = E_x\epsilon/(1-\nu)$. We obtain \SI{472}{\mega\pascal} for the $0^\circ$ direction.

\begin{figure}[b!htp]
    \centering\includegraphics{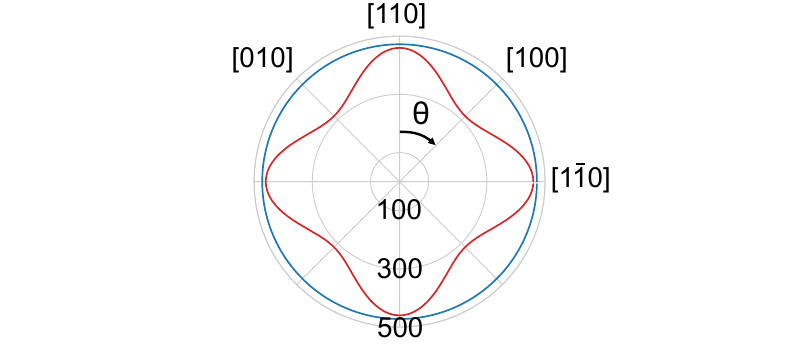}
    \caption{As-grown stress, $\sigma_{\mathrm{grown}}$, is in blue solid line, and released stress is in red (in MPa).}
    \label{fig:stress}
\end{figure}

The released stress in the string resonators can be obtained by taking into account Poisson's ratio, $\sigma_{\mathrm{grown}}(1-\nu)$, resulting in \SI{460}{\mega\pascal} and \SI{319}{\mega\pascal} for a string oriented along 0$^\circ$ and 45$^\circ$, respectively, see \fref{fig:stress}.

\subsection{Optical properties}

\begin{figure}[t!hbp]
    \centering\includegraphics{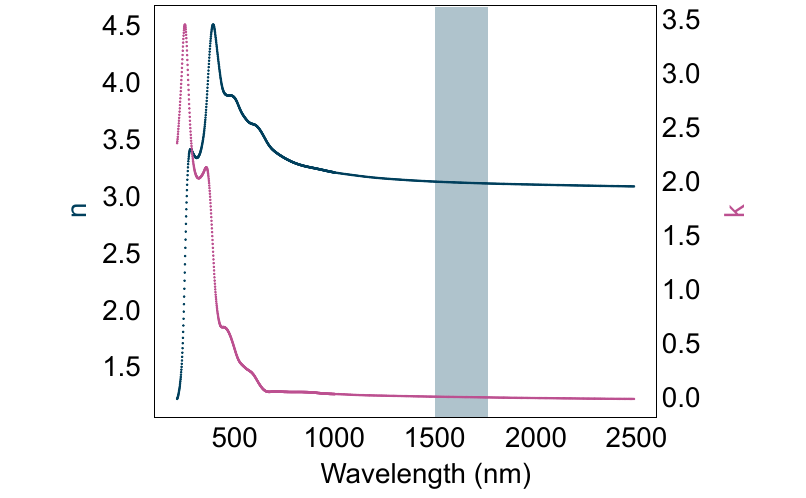}
    \caption{Refractive index $n$ and extinction coefficient $k$ for the \SI{73}{\nano\meter}-thick InGaP film determined from ellipsometer measurements. The refractive index $n$ is in agreement with Ref.~\cite{adachi1982refractive}.}
    \label{fig:ellipsometer}
\end{figure}

We measured the refractive index and absorption coefficient of the \SI{73}{\nano\meter}-thick \InGaPval{} layer using an ellipsometer. The results are shown in \fref{fig:ellipsometer} and resemble the values found in the literature for the wavelength range of interest in our work \cite{adachi1982refractive}. 

\section{Mechanical damping}
\label{app:damping}

The total mechanical quality factor of a mechanical resonator is given by \cite{imbodenDissipationNanoelectromechanicalSystems2014,schmid2016fundamentals}

\begin{equation}
    Q^{-1} = Q_{\mathrm{int}}^{-1} + Q_{\mathrm{ext}}^{-1} = \sum_i \left(Q_{\mathrm{int}}^{i}\right)^{-1}+\sum_j \left(Q_{\mathrm{ext}}^{j}\right)^{-1},
\end{equation}
where $Q_{\mathrm{int}}^{i}$ originates, amongst others, from surface loss, thermoelastic damping, or loss processes related to material defects, and $Q_{\mathrm{ext}}^{i}$ from, amongst others, gas damping or clamping loss. For an in-depth discussion, see Refs.~\cite{imbodenDissipationNanoelectromechanicalSystems2014,schmid2016fundamentals}. In the following, we analyze mechanical loss processes relevant to this work.

\subsection{Clamping loss}

Clamping loss is determined by the geometry of the clamping of the mechanical resonator to the substrate. Clamping results in the transfer of acoustic energy from the mechanical resonator to its environment. In the following, we will estimate the clamping loss for string-type resonators and compare it to the one of a cantilever and a fully clamped square membrane which would give the lower and upper bounds for clamping loss, respectively. Further, we will examine the clamping loss of a fully clamped membrane to obtain an upper limit to the clamping loss of a trampoline resonator, which we also independently simulated with FEM. 

For string resonators in this work, we can write the total mechanical quality factor as
\begin{equation}\label{eq:stringclampQ}
    Q^{-1}_{\mathrm{total}} = \frac{1}{D_n \cdot Q_{\mathrm{int}}} + \frac{1}{Q_\textrm{clamp,s}}.
\end{equation}
Similar to Ref.~\cite{schmidDampingMechanismsHigh2011}, we assume  $Q_\textrm{clamp,s} = \eta L/h$, where $\eta$ is a free parameter. For the string resonators in our work  and using \eref{eq:stringclampQ}, we obtain $Q_\textrm{clamp,s} \sim 10^{12}$. 

We can compare the clamping loss of the doubly-clamped string resonators to the one of singly-clamped cantilevers, where the latter will give an upper bound on $Q_\textrm{clamp}$ for string resonators. The clamping loss of a cantilever of thickness $h$, width $w$ and length $L$ is given by \cite{schmid2016fundamentals}
\begin{equation}\label{eq:clamping_cantilever}
    Q^{-1}_{\mathrm{clamp,c}}  = 0.31 \frac{w}{L}\left(\frac{h}{L}\right)^4,
\end{equation}
for the case where the thickness of the substrate is comparable to the wavelength of the acoustic wave. The fundamental mode of a \SI{155}{\micro\meter}-long and  \SI{2}{\micro\meter}-wide cantilever has $Q_\textrm{clamp,c}=5 \cdot 10^{15}$, which is larger than $Q_\textrm{clamp,s}$ of a string with the same dimensions. 

Similarly, we obtain the lower bound on $Q_\textrm{clamp,s}$  by examining a worst-case scenario of a fully clamped membrane. The clamping loss-limited $Q$ factor of a tensile-stressed square membrane clamped to a thick substrate is given by \cite{villanueva2014evidence,schmid2016fundamentals}
\begin{equation}
    Q_\textrm{clamp,m} \approx \frac{3}{2} \sqrt{\frac{\rho_\textrm{InGaP}}{\rho_\textrm{GaAs}}} \left( \frac{E_\textrm{GaAs}}{\sigma}\right)^{3/2}\frac{n^2 m^2}{(n^2 + m^2)^{3/2}}\frac{L}{h},
    \label{eq:clamping_membrane}
\end{equation}
where $h$ is the thickness of the membrane, $L$ the length of one of its sides, $\rho_\textrm{InGaP}$ and $\rho_\textrm{GaAs}$ are the densities of the membrane and the substrate, respectively, $E_\textrm{GaAs}$ is Young's modulus of the GaAs substrate, $n$ and $m$ are the eigenmode numbers of the mechanical mode. For a $155 \times $\SI{155}{\micro\meter}$^2$ square membrane with a uniform tensile stress of \SI{415}{\mega\pascal}, $Q_\textrm{clamp,m}$ for the fundamental mode is about $7 \cdot 10^6$. As we observe $Q$ factors of string resonators  $\leq 3\cdot 10^6$, we conclude that they are not limited by clamping loss.

Now, we estimate a lower bound on $Q_\textrm{clamp}$ for the trampoline resonators by using \eref{eq:clamping_membrane}. For a $1130 \times $\SI{1130}{\micro\meter}$^2$ square membrane (corresponding to a trampoline with tether length of \SI{750}{\micro\meter} and central pad size of \SI{100}{\micro\meter}) with uniform tensile stress of \SI{415}{\mega\pascal}, $Q_\textrm{clamp,m}$ for the fundamental mode is about $5 \cdot 10^7$.  This value yields a lower limit for the clamping loss-related $Q$ factor of trampoline resonators. We also performed FEM simulations to estimate $Q_\textrm{clamp,m}$ for the trampoline resonators using perfectly matched layers \cite{romero2020engineering}. We find that $Q_\textrm{clamp}>10^8$ \cite{fia2022tensile}. Therefore, we consider that the micromechanical resonators of this work are not clamping-loss limited.




\subsection{Gas damping}

Gas damping is another extrinsic mechanical damping process \cite{imbodenDissipationNanoelectromechanicalSystems2014}. For a mechanical resonator, such as a string- or membrane-type resonator, oscillating at frequency $\Omega_m$ at low pressures $P$, i.e., in the ballistic regime, where the mean free path of the gas molecules is larger than the dimensions of the mechanical resonator, $Q_\textrm{gas}$ is given as \cite{verbridge2008size}

\begin{equation}
    Q^{-1}_\textrm{gas} = 4\sqrt{\frac{2}{\pi}}\frac{1}{\rho h \Omega_m}P\sqrt{\frac{M}{RT}},
    \label{eq:q_gas}
\end{equation}
where $M$ is the molecular mass of the gas molecules, $R$ is the molar gas constant, and $T$ is the temperature of the gas.

\begin{figure}[t!hbp]
    \centering\includegraphics[width=0.95\columnwidth]{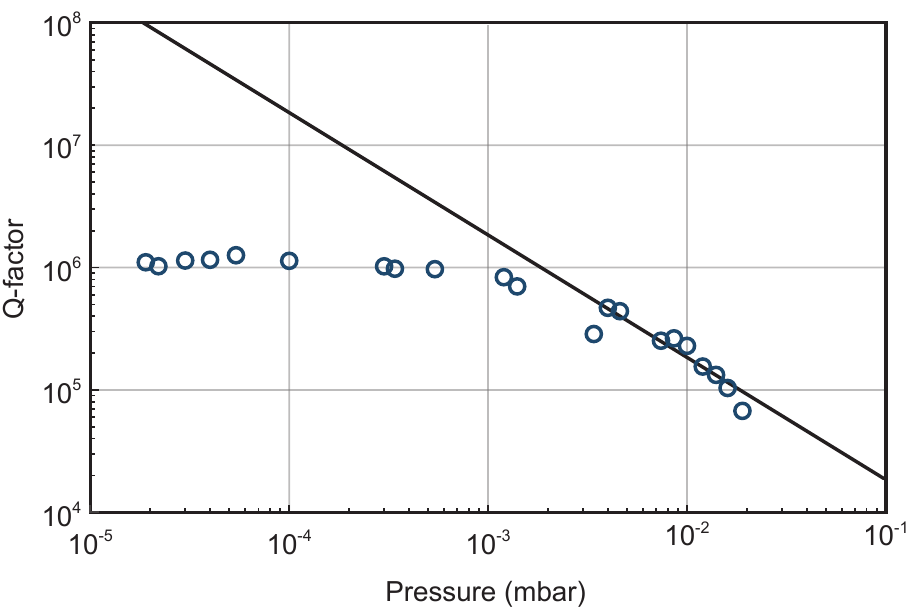}
    \caption{Pressure-dependent measurement of the $Q$ factor of InGaP string resonators. The predicted gas damping from \eref{eq:q_gas} is shown as solid line.}
    \label{fig:QvsPstrings}
\end{figure}

\fref{fig:QvsPstrings} shows pressure-dependent measurements for InGaP string resonators of length \SI{155}{\micro\meter} and width \SI{2}{\micro\meter} oriented along $0^{\circ}$. For pressures $P>10^{-2}\,$mbar, the $Q$ factor of the strings is limited by gas damping, while for pressures $P<10^{-2}\,$mbar, $Q$ is constant, indicating another limiting damping mechanism.

\subsection{Thermoelastic damping}

Thermoelastic damping (TED) is related to the irreversible conversion of mechanical energy into heat \cite{imbodenDissipationNanoelectromechanicalSystems2014}. During the oscillation of the mechanical resonator, one of its sides is compressed, while the opposite side is stretched. This deformation produces a temperature gradient and, thus, an irreversible transport of heat, in case the thermal expansion coefficient $\alpha$ is non-zero.

\begin{figure}[t!hbp]
    \centering\includegraphics{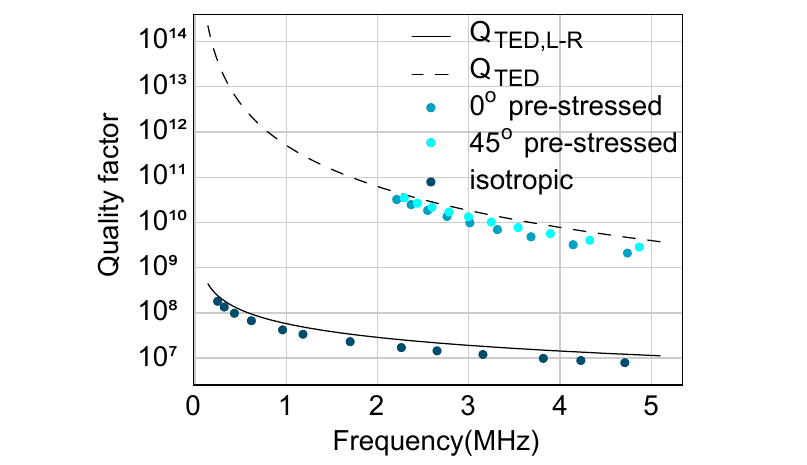}
    \caption{TED loss in beams. Analytical  and FEM models of $Q_\textrm{TED}$ for \SI{2}{\micro\meter}-wide, \SI{73}{\nano\meter}-thick InGaP beams. Solid line is \eref{eq:ted-l-r}, dashed line is \eref{eq:ted-l-r-s} and dots are FEM results.}
    \label{fig:ted_beam}
\end{figure}

For a thin beam of length $L$, thickness $h$, and width $w$ at temperature $T$ the quality factor limited by TED, $Q_\textrm{TED}$, is given by \cite{lifshitz2000thermoelastic} 

\begin{equation}
    Q^{-1}_\textrm{TED, L-R} = \frac{\alpha^2 E T}{C_p}
    \left( \frac{6}{\xi^2} - \frac{6}{\xi^3}\frac{\sinh\xi + \sin\xi}{\cosh\xi + \cos\xi}\right),
    \label{eq:ted-l-r}
\end{equation}
where $\xi = h\sqrt{\frac{\omega_0}{2\chi}}$ is a dimensionless variable, with $\omega_0$ the isothermal value of the eigenfrequency $\omega$, $\chi = \kappa/\rho C_p$  the thermal diffusivity, $C_p$  the heat capacity
per unit volume at constant pressure, and $\kappa$ the thermal conductivity. At room temperature, the linear thermal expansion coefficient is $\alpha \approx 4.28 \times 10^{-6}$ 1/K, based on the experimental values for InP and GaP \cite{kagaya1986mode}.
For this isotropic model, we set Young's modulus to $E = 85$\,\SI{}{\giga\pascal}. Based on \eref{eq:ted-l-r}, we  obtain the straight line in \fref{fig:ted_beam} for $Q_\textrm{TED, L-R}$. We independently performed FEM simulations of beams with the same dimensions and material parameters and obtain very similar results (see \fref{fig:ted_beam}).


As our InGaP resonators are strained we need to modify the thermal relaxation time and the oscillation frequency in the model above \cite{zener1938internal}. Following the derivation from Ref.~\cite{kumar2010stress} the pre-stress is included via an additional axial force $\sigma_0 = \frac{F}{wh}$, which leads to $Q_\textrm{TED}$ for strained beams
\begin{equation}
    Q^{-1}_\textrm{TED} = \frac{1}{1 + a\frac{F}{F_\textrm{cr}}}\cdot Q^{-1}_\textrm{TED, L-R}(\xi),
    \label{eq:ted-l-r-s}
\end{equation}
where $F_\textrm{cr} = \pi EI/L^2$, $a = 0.97$ is a factor that depends on the boundary condition \cite{bokaian1990natural}, $I$ is the moment of inertia, and $\xi$ gets modified by inserting $\omega_0 = \frac{\pi}{L^2}\sqrt{\frac{EI}{\rho w h}}\sqrt{\pi^2 + \frac{FL^2}{EI}}$.

For a stress of $\sigma_0 = \SI{300}{\mega\pascal}$, $Q_\textrm{TED}$ of a strained beam is shown as the dashed line in \fref{fig:ted_beam} and considerably larger than $Q_\textrm{TED, L-R}$. We also simulated such stressed beams in FEM and obtain a good agreement with the analytical estimate \eref{eq:ted-l-r-s} (see \fref{fig:ted_beam}).

Estimating TED of the trampoline patterned with a PhC should take into account the periodic perforation of the material, which leads to an effective medium with a weaker elasticity matrix \cite{rabinovich1997effect} and periodic stress redistribution. The periodic strain variation results in a temperature gradient around the PhC hole edges and, therefore, in an increase of TED loss \cite{tu2012study}. 

To get an estimate of $Q_\textrm{TED}$ for a PhC trampoline resonator, we performed FEM simulations (using $r_\textrm{PhC} = $\SI{552}{\nano\meter}, $a_\textrm{PhC} =$\SI{1323}{\nano\meter}) and obtained $Q_\textrm{TED} \approx 10^8$ for the fundamental mode at a frequency of $\Omega_m/2\pi = 163$\,\SI{}{\kilo\hertz}, which is higher than the $Q$ we observed in the experiment. 


\subsection{Dilution factor in FEM}

$Q_{\textrm{int}}$ can be generally written as the ratio of the stored versus the lost energy over one cycle of oscillation \cite{schmid2016fundamentals}:

\begin{equation}
    Q_{\textrm{int}} = 2\,\pi \frac{W_{\textrm{elongation}} + W_{\textrm{bending}}}{\Delta W_{\textrm{elongation}} + \Delta W_{\textrm{bending}}},
\end{equation}
where $W_{\textrm{elongation}}$ and $W_{\textrm{bending}}$ are the potential energy stored in the elongation and bending of the resonator, respectively. Similarly, $\Delta W_{\textrm{elongation}}$ and $\Delta W_{\textrm{bending}}$ are the corresponding loss processes.

The quality factor can be increased by diluting the intrinsic material friction. This can be achieved by utilizing materials with high intrinsic tensile stress, which introduces $W_\textrm{tensile}$ as an additional contribution to the potential energy. The increased quality factor $ Q_{\textrm{D}}$ is then given by \cite{Federov2019,sementilli_nanomechanical_2022} 
\begin{equation}
    Q_{\textrm{D}} = D \cdot  Q_{\textrm{int}},    
\end{equation}
where $D$ is the dilution factor that depends on the pre-strain of the material, resonator geometry, and the displacement mode profile. The dilution factor is in general given by \cite{sementilli_nanomechanical_2022}
\begin{equation}
    D = 1 + \frac{W_{\textrm{tensile}}}{W_{\textrm{elongation}} + W_{\textrm{bending}}}.
    \label{eq:dil}
\end{equation}
We see that by increasing the elastic energy $W_{\textrm{tensile}}$, the mechanical quality factor of the resonator can be increased to a considerable extent, with record values of $Q>10^{10}$ in highly tensile-strained crystalline silicon \cite{beccari2022strained}.

For thin two-dimensional membranes (i.e., membranes with a large length-to-thickness ratio) with the out-of-plane displacement $u_z = w(x,y)$, the dilution factor can be calculated using FEM. In the case of $D_Q \gg 1$, the first term in \eref{eq:dil} can be neglected. Then, the dissipation dilution factor can be calculated by integrating over the entire membrane area as \cite{Fedorov_Thesis}

\begin{widetext}
    \begin{equation}
        D_Q = \frac{\rho h\Omega_m^2}{D_p}\frac{\displaystyle \iint w(x,y)^2dxdy}{\displaystyle \iint \biggl\{ \left( \frac{\partial^2 w(x,y)}{\partial x^2} + \frac{\partial^2 w(x,y)}{\partial y^2} \right)^2 + 2(1-\nu)\left[ \left( \frac{\partial^2 w(x,y)}{\partial x \partial y}  \right)^2 - \frac{\partial^2 w(x,y)}{\partial x^2} \frac{\partial^2 w(x,y)}{\partial y^2} \right] \biggl\} dxdy}
    \end{equation}
\end{widetext}
where $D_p = Eh^3/[12(1-\nu^2)]$ is the flexural rigidity of the material and $\nu$ is Poisson's ratio. The numerator yields the kinetic energy of the membrane and the denominator the energy stored in the membrane bending \cite{landau1986theory}.


\section{Experimental parameters}
\label{app:params}

\tref{tab:params} summarizes the material and device parameters of this work.
\begin{table}[t!hbp]
    \centering
        \begin{tabular}{l c c}
        \hline\hline
        Parameter   & Symbol    & Value\\\hline
        \multicolumn{3}{l}{\textbf{Material properties}}\\\hline
        Ga content  & $x$   & 0.5658 \\\hline
        Density (\SI{}{\kilo\gram}/\SI{}{\meter}$^3$)   & $\rho$    &  $(4.81 - 0.67x) \cdot 10^3$\\\hline
        Thickness (\SI{}{\nano\meter}) & $h, d_\textrm{PhC}$    & 73 \\\hline
        Stress (\SI{}{\mega\pascal}) & $\sigma(0^{\circ})$  & $467.7 \pm 7.1$
        \\ & $\sigma(45^{\circ})$   & $313.3 \pm 5.4$
        \\ & $\sigma(90^{\circ})$   & $374.9 \pm 16.4$
        \\ \hline
        Elastic constants (\SI{}{\pascal}) & $c_{11}$   &  $(10.11 + 3.94x) \cdot 10^{10}$ \\
         at \SI{300}{\kelvin}  & $c_{12}$   &  $(5.61 + 0.59x) \cdot 10^{10}$  \\
          &  $c_{44}$   &  $(4.56 + 2.47x) \cdot 10^{10}$ \\ \hline
        Lattice constant \cite{shur1996handbook}  (\SI{}{\angstrom})& & \\
        \InGaPx & $a_{\textrm{\InGaPx{}}}$ & $(5.8687-0.4182x) $\\
        GaAs & $a_\textrm{GaAs}$ & $5.65325 $\\\hline
        Thermal diffusivity (\SI{}{\watt/\kelvin}) & $\kappa$ & $66$ \\\hline
        Heat capacity (\SI{}{\joule/\kilo\gram\kelvin}) & $C_p$ & $380$ \\\hline
        Linear thermal expansion & $\alpha$ & $4.28 \cdot 10^{-6}$ \\
        coefficient at RT (\SI{}{1/\kelvin})& & \\\hline
        \multicolumn{3}{l}{\textbf{Optical properties}}\\\hline
        Refractive index & $n$ & 3.15 \\
        \hline
        Gap (\SI{}{\micro\meter}) & $L_\textrm{gap}$ & 14.8\\\hline
        \multicolumn{3}{l}{\textbf{Mechanical properties}}\\\hline
        Effective mass (\SI{}{\nano\gram})& $m_\textrm{eff}$& 9.3\\\hline 
        \hline\hline
        \end{tabular}
    \caption{Parameters used in FEM and RCWA simulations.}
    \label{tab:params}
\end{table}




\section{InGaP mechanical resonators for force sensing and integrated microcavity optomechanics}
\label{app:appsfuture}

We have demonstrated InGaP trampoline mechanical resonators with a quality factor of $10^7$ at room temperature with a resonance frequency of $38\,$kHz, resulting in a $Q\cdot f_m$ product of $7\cdot 10^{11}\,$Hz.

In the following, we discuss future device improvements and mention applications, where improved InGaP trampoline resonators are particularly fitting. We focus our discussion on InGaP mechanical resonators that offer a central area amenable to patterning with a photonic crystal (PhC). We deliberately choose the use of a suspended PhC as engineering the reflectance of the mechanical resonator via the PhC enables efficient transduction of mechanical displacement to out-of-plane light fields, as was recently demonstrated \cite{zhouCavityOptomechanicalBistability2022}. Furthermore, the  (Al,Ga)As combined with InGaP material platform provides the opportunity to realize integrated free-space optomechanical microcavities \cite{Kini_Thesis}. Another interesting avenue would be to  pattern phononic-shielded string-like resonators in InGaP \cite{bereyhi2022hierarchical,Bereyhi2022Perimeter}, which we will not discuss in the following.

\paragraph{Improvements of mechanical performance}

Figures of merit in nanomechanics are the mechanical frequency $f_m$, the mechanical $Q$ factor, and the effective mass $m_\textrm{eff}$ of the resonator. An increase in $Q$ and $f_m$ can be achieved by increasing the Ga content of InGaP to obtain a tensile stress of up to 1\,GPa. This is a factor of two higher than the stress in the current samples and would translate into a factor of about $\sqrt{2}$ enhancement of $f_m$. In general, the tensile strain of the InGaP layer allows employing more advanced strain engineering techniques than we have currently used. This comprises optimized trampoline designs (with $Q$ factors up to $10^8$ at $f_m\sim150\,$kHz in SiN \cite{norte2016mechanical}), hierarchical clamping ($Q$ factor larger than $10^8$ at $f_m\sim100\,$kHz  in SiN \cite{bereyhi2022hierarchical}), or phononic-shield membranes ($Q$-factor values larger than $10^8$ at $f_m\sim700\,$kHz in SiN \cite{tsaturyan2017ultracoherent}). Furthermore, decreasing the thickness of the InGaP layer by a factor of two to about 35\,nm, would lead to a further increase of $Q$, with a concomitant decrease of $f_m$. 

Overall, we can expect to reach $Q$ factors of InGaP trampoline-like resonators larger than $10^8$ at frequencies of between $50$ to $100$\,kHz with an effective mass between 1 to 10\,ng. Furthermore, we expect that operating InGaP resonators at low temperatures will result in a further enhancement of $Q$, as observed in other crystalline materials such as diamond \cite{taoSinglecrystalDiamondNanomechanical2014} or silicon \cite{taoSinglecrystalDiamondNanomechanical2014,beccari2022strained}, with an increase of up to a factor of five.  

\paragraph{Force sensing}
For our current InGaP trampolines, we estimate a thermal noise limited force sensitivity of $F_\mathrm{th}\approx 50$\,aN/$\sqrt{\mathrm{Hz}}$ (for an effective mass of $10\,$ng). Improved InGaP trampoline resonators with an effective mass of 1\,ng at a frequency of $50\,$kHz and a $Q$ of $10^8$ would yield a thermal noise limited force sensitivity at room temperature of 7\,aN/$\sqrt{\mathrm{Hz}}$, which would be better than the one achieved with SiN trampolines (19.5\,aN/$\sqrt{\mathrm{Hz}}$ \cite{norte2016mechanical}) or phononic-shield membranes (37\,aN/$\sqrt{\mathrm{Hz}}$ \cite{tsaturyan2017ultracoherent,halgMembraneBasedScanningForce2021}) and similar to other mechanical resonator geometries designed for force sensing \cite{heritierNanoladderCantileversMade2018,debonisUltrasensitiveDisplacementNoise2018,sahafiUltralowDissipationPatterned2020}. When operating the devices at lower temperatures, e.g., at $4\,$K, we estimate $0.37$\,aN/$\sqrt{\mathrm{Hz}}$ when also assuming a $Q$ increase by a factor of 5 upon cooling. This value is close to the sensitivity required to detect the magnetic moment of a single proton \cite{taoUltrasensitiveMechanicalDetection2016}.
    
\paragraph{Quantum optomechanics at room temperature}

Improved InGaP trampoline resonators with a frequency of $100\,$kHz and a $Q$ of $10^8$ would yield a $Q\cdot f\sim 10\cdot10^{12}\,$Hz, which places such devices in the regime of quantum optomechanics at room temperature. As the trampoline can be patterned with a high-reflectivity PhC, it can be directly used as a mirror in an optical cavity, which enables feedback cooling of a trampoline mechanical mode to the ground state \cite{rossiMeasurementbasedQuantumControl2018,magriniRealtimeOptimalQuantum2021}.
    
\paragraph{Optomechanical microcavity}

We have recently demonstrated an optomechanical microcavity based on an (Al,Ga)As heterostructure \cite{Kini_Thesis}, which accesses the regime of ultra-strong optomechanical coupling, with $\omega_m\sim0.3 g_0$. However, these experiments were limited by the large optical loss rate of hundreds of GHz, placing the system deep in the bad cavity regime, and the low mechanical $Q\sim10^4$ of the suspended GaAs-based mechanical resonator. 

Using InGaP trampolines for such an integrated optomechanical microcavity would allow reaching $Q$ factors that are four orders of magnitude larger compared to GaAs-based mechanical resonators. Furthermore, the strained InGaP layer enables patterning of a PhC on a trampoline with a sufficiently large central pad of \SI{250}{\micro\meter}$\times$\SI{250}{\micro\meter}, such that diffraction loss and collimation or finite size effects \cite{toftvandborgCollimationFinitesizeEffects2021} of the PhC when illuminated with a \SI{50}{\micro\meter} waist are estimated to be negligible. A reflectivity above $0.9998$ has been recently demonstrated with a similarly sized suspended PhC membrane at telecom wavelength \cite{zhouCavityOptomechanicalBistability2022}, yielding a Finesse $\mathcal{F}>3\cdot10^4$. 

An envisioned InGaP-based optomechanical microcavity with a length $L_c\sim \lambda/2$ at $\lambda_0=1550\,$nm, would yield an optical loss rate  $\kappa=\pi c/(2L_c\mathcal{F})< 2\pi \cdot 3\,$GHz. Thanks to the short length of $\lambda_0/2$ of the microcavity a single-photon coupling strength $g_0\sim2\pi\cdot 0.9\,$MHz is within reach. This optomechanical microcavity would be placed in the ultra-strong coupling ($g_0/\omega_m\sim 9$ with $\omega_m=2\pi\cdot 100\,$kHz) and bad cavity regime ($g_0/\kappa\sim 3\cdot 10^{-4}$). With a $Q$ of $10^8$ ($5\cdot 10^8$) at room temperature ($4\,$K), a single-photon cooperativity $C_0=(g_0)^2/\kappa\gamma n_{th}$ of $4\cdot 10^{-3}$ ($1.6$) can be achieved, with the phonon occupation $n_{th}=k_BT/\hbar\omega_m$. Such a system will allow studying novel optomechanical effects in the ultra-strong coupling regime \cite{friskkockumUltrastrongCouplingLight2019} and realization of mechanical squeezing as recently proposed in Ref.~\cite{kusturaMechanicalSqueezingUnstable2022}.

An integrated optomechanical microcavity may also be able to access the single-photon strong coupling regime, which is so far elusive in nano- or micromechanics. One approach relies on multi-element optomechanics \cite{xuerebStrongCouplingLongRange2012}, which  aims at increasing $g_0$ by virtue of collective mechanical effects between highly reflective mechanical elements placed within a cavity. We have taken steps in this direction in the (Al,Ga)As-based system by realizing two-element mechanics on a chip \cite{Kini_Thesis}. Using InGaP instead of GaAs-based mechanics would thereby allow reaching higher $Q$ mechanical resonators. Further, a key challenge with this approach is to match the mechanical frequency of the resonators within their linewidth. As InGaP is a piezoelectric material, one can tune the mechanical frequency in-situ via application of a static voltage, which is a desired feature to realize the stringent requirements of Ref.~\cite{xuerebStrongCouplingLongRange2012}.

Another approach to reach the single-photon strong coupling regime relies on minimizing the optical loss rate $\kappa$ via the use of photonic bound states in the continuum \cite{fitzgeraldCavityOptomechanicsPhotonic2021}. A system composed of two InGaP-based suspended PhC mirrors at a well-defined spacing, e.g., defined via heterostructure growth, can realize such a photonic bound states in the continuum, where it has to be seen in the experiment how small $\kappa$ can be realized given fabrication imperfections of the PhC and the material's absorption.

\clearpage

\bibliography{InGaP_bib}

\end{document}